\documentclass[10pt]{revtex4-1}

\usepackage[english]{babel}
\usepackage[utf8]{inputenc}
\usepackage{graphicx}
\usepackage{amsmath}
\usepackage{amsfonts}
\usepackage{microtype}
\usepackage{siunitx}
\usepackage{multirow}

\graphicspath{{./}{./images/}}

\begin{document}
\title{Critical threshold for microtubule amplification through templated severing}
\author{Marco Saltini}
\author{Bela M. Mulder}
\affiliation{AMOLF, Science Park 104 1098XG Amsterdam, Netherlands}
%\date[Date: ]{September, 2018}
%\revised[Revised: ]{\today}
\begin{abstract}
%We present a one dimensional model of dynamic microtubules with nucleation of new filaments due to a well-established biological mechanism that consists of the severing of microtubules at specific positions along their length, i.e. where they cross differently oriented microtubules. Simultaneously, every severing of a microtubule shortens its length and amplifies their number. We show that the behaviour of the newly-created plus end after severing (i.e. whether the dynamic tip of a microtubule grows or shrink after a severing event) strongly influences both the speed of such an amplification and its success probability. Then, we identify a critical relationship between such a success probability and the probability that a newly-created plus end after severing grows, provided that the overall depolymerization speed of the microtubules outweighs the polymerization one. Next, we show how this critical relationship can be quantified using a combination of computer simulations and a new analytical technique to calculate the first passage time distribution for a microtubule to reach a target. Finally, we approximate the model by reducing the number of possible severing spots to two, and we demonstrate that such an approximation describes the entire system with an high degree of accuracy.
The cortical microtubule array of dark-grown hypocotyl cells of plant seedlings undergoes a striking, and developmentally significant, reorientation upon exposure to light. This process is driven by the exponential amplification of a population of longitudinal microtubules, created by severing events localized at crossovers with the microtubules of the pre-existing transverse array. We present a dynamic one-dimensional model for microtubule amplification through this type of templated severing. We focus on the role of the probability of immediate rescue-after-severing of the newly-created lagging microtubule, observed to be a characteristic feature of the reorientation process. Employing stochastic simulations, we show that in the dynamic regime of unbounded microtubule growth, a finite value of this probability is not required for amplification to occur, but does strongly influence the degree of amplification, and hence the speed of the reorientation process. In contrast, in the regime of bounded microtubule growth, we show that amplification only occurs above a critical threshold. We construct an approximate analytical theory, based on a priori limiting the number of crossover events considered, which allows us to predict the observed critical value of the rescue-after-severing probability with reasonable accuracy. 
\end{abstract}
\maketitle

\section{Introduction}

Microtubules are a ubiquitous component of the cytoskeleton in eukaryotic cells. They are filamentous aggregates of tubulin-dimers reaching lengths of several $\mu$m's. They are typically part of structures that span the dimensions of the whole cell, enabling e.g.\ their major role in intracellular transport by providing ``tracks'' for cargo carrying motor proteins and cell division where they form the mitotic spindle responsible for the spatial segregation of the duplicated chromosomes \cite{Alberts2007}. The fact that during the cell cycle microtubules can be reassembled into different spatial structures is due to their intrinsically dynamic nature. They stochastically switch between phases of growth through polymerization to phases of shrinkage through rapid depolymerization, a mechanism that has been dubbed dynamic instability \cite{Mitchison1984DynamicGrowth}. A direct consequence of this mechanism is that microtubules have a finite lifetime, as they can shrink away, and therefore need to be actively (re)nucleated to sustain their overall number. Cells achieve control over the microtubule structures they build by manipulating the nucleation and dynamics of microtubules in space and time, using specific nucleating complexes and a host of microtubule-interacting proteins (MAPS)\cite{Sedbrook2004MAPsOrganization}.

%Microtubules are polarized polymers built by tubulin dimers, and they are a component of the eukaryotic cellular cytoskeleton. They are very important for the dynamics of the cell, indeed, e.g. they form the mitotic spindle, enable cell migration, or act as pathways for motor proteins allowing them to transport cargoes \cite{Alberts2007}. In order to perform all these operations, microtubules have an intrinsic dynamic instability mechanism, that makes them dynamically switch from growing (polymerizing) to shrinking (depolymerizing) state, and vice versa \cite{Mitchison1984DynamicGrowth}.  

Growing plant cells have a unique microtubule structure called the cortical array. The cortical array is an assembly of mutually aligned microtubules localized close to the cell membrane that almost homogeneously covers the inside surface of the cell. Generically the preferential direction of the cortical microtubules is transverse to the long axis of the cell. This is crucial to their function, as they guide the anisotropic deposition of cell-wall building polymers, which in turn allows the cell to grow along a single expansion axis. In this way the cortical array drives the dominant mode of morphogenesis in plants, which is the formation of linear extensions, like roots, stems and branches. However, it is known that this generic growth scenario can be modulated by hormonal, mechanical and other environmental signals \cite{Dixit2004TheOrganization.}. A striking example of this type of modulation is the reorientation of the cortical array of dark-grown hypocotyl (stem precursor) cells after exposure to blue light \cite{Lindeboom2013}. This effect is highly relevant, as the developmental program of the plant must change dramatically, once the hypocotyl, which typically emerges from a buried seed, first reaches the sunlight. It is believed that the observed reorientation from the transverse to the longitudinal orientation of the cortical array is associated with the arrest of further growth, and the subsequent differentiation of the cells. 

The light-induced reorientation of the cortical array is mediated by the microtubule severing protein \emph{katanin}. It has been shown to localize at the crossover between differently oriented cortical microtubules and, moreover, to preferentially sever the overlying microtubule, i.e.\ the one that crossed over a pre-existing one. As the underlying microtubule is most likely to be a transversely oriented microtubule from the pre-exposure state and severing effectively multiplies the number of microtubules, this effect can rapidly create an exponentially growing population of longitudinal microtubules. In this way the original transverse cortical array serves as a template for the reorientation towards a longitudinal array. 

Recent experiments involving a number of mutants in which the reorientation effect is impaired, have shown that there is an important role for the propensity of the newly-created plus end of the lagging microtubule to immediately switch to the growing state, a process which \emph{in vivo} is mediated by the prominent MAP \emph{CLASP} \cite{Nakamura2018}. This specific function of \emph{CLASP} appears to have evolved, as the default outcome of a severing event is the creation of a shrinking plus end of the lagging microtubule \cite{Tran1997AEnds}. Stochastic simulations of a simplified model of the reorientation mechanism, indeed, showed that the degree of amplification of the numbers of microtubules due to severing increases monotonically with the probability of the so-called rescue-after-severing of the lagging plus end \cite{Nakamura2018}. However, the dynamic parameters of the microtubules measured in the experiments suggest that, at least in the initial phase of reorientation, the microtubules are in the so-called unbounded-growth regime \cite{Dogterom1993PhysicalStructures}. Since in this regime the microtubules in principle are very long-lived on the timescale of the reorientation, this raises the question to what extent rescue-after-severing is in fact a necessary ingredient of the mechanism. Moreover, as the total amount of tubulin in the cell is finite, it is also clear that unbounded-growth and amplification of microtubules cannot be sustained indefinitely, as the pool of available tubulin to drive polymerization is inevitably depleted. This will cause the growth speed of the microtubules to decrease, effectively driving them back to the bounded-growth regime. To fully understand the reorientation process we thus need to disentangle the role of the microtubule growth state from that of the probability of rescue-after-severing, and this is the main aim of this paper. We approach this problem using a combination of stochastic simulations and analytical theory, which together allow us to fully characterize the requirements for the amplification of longitudinal microtubules to occur.

The structure of the paper is as follows. In Sec.\ II we introduce our dynamic model of longitudinal microtubules undergoing dynamic instability and severing in the presence of a grid of stable transverse microtubules. We then briefly review some of the main features of Dogterom-Leibler model for microtubules undergoing dynamic instability \cite{Dogterom1993PhysicalStructures}, on which our model is based. In Sec.\ III we show how, for microtubules in the unbounded-growth regime, while not a necessary ingredient, the probability of rescue-after-severing does dramatically affect the speed and ultimate success probability of amplification of the longitudinal microtubule population. Then, we extend our treatment to the bounded-growth regime, showing that in this case amplification can occur provided that the probability of rescue-after-severing exceeds a critical threshold. Finally, we calculate the critical value for the probability of rescue-after-severing using a combination of analytical calculations and computer simulations. To that end, we introduce an approximate theory in which each microtubule can experience at most two crossovers, allowing an analytical determination of the contribution of the probability of rescue-after-severing to the probability that the creation of a crossover actually leads to a severing event, and hence contributes to the amplification. In order to do so, we develop a novel approximate technique to calculate the first-passage time distribution (hereafter FPTD) for the microtubules to reach relatively close targets, which has potential application for studying first-passage time problems in other systems as well.

\section{The model}

\subsection{Dynamic model}

\begin{figure}[htbp]   
  \centering
  \includegraphics[scale=0.27]{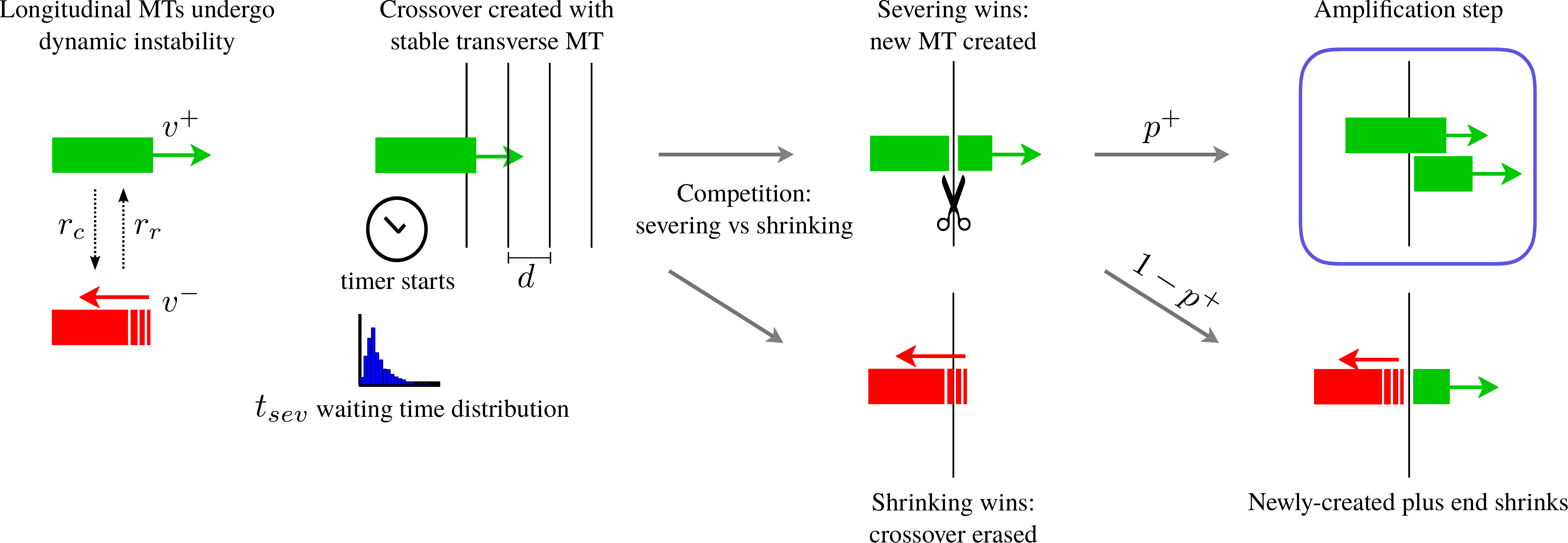}
  \caption{Schematic of the model of longitudinal microtubules undergoing dynamic instability in a grid of stable transverse microtubules. After a crossover creation, a competition between the intrinsic severing waiting time and crossover removal due to dynamic instability takes place. If severing occurs, the newly-created plus end is rescued with probability $p^+$.}  \label{fig1_model}
\end{figure}

In order to better understand the importance of the probability of rescue-after-severing for the reorganization mechanism of the cortical microtubule array, we introduce a stochastic model of longitudinal microtubules undergoing dynamic instability in a one dimensional grid of transverse microtubules. Since before the exposure to light the transverse array consists of relatively long microtubules, in the initial stage of this process, i.e. the first $500$ seconds, the array can be seen as a constant background. In particular, we focus on this time interval because it is the period of time in which the amplification takes place.

The model consists of a single longitudinal microtubule undergoing dynamic instability in the one dimensional grid of stable, transverse microtubules with constant spacing $d$ between neighboring filaments \cite{Nakamura2018}. According to experimental observations, where the angle between differently oriented microtubules is very close to $\ang{90}$, we assume that all longitudinal microtubules are exactly perpendicular to the transverse. The microtubule is nucleated at position $x=0$ with plus end in the growing state with growing speed $v^+$, and it can switch to shrinking state with constant catastrophe rate $r_c$. Once it is in the shrinking state, either its plus end shrinks back to position $x=0$ and dies, or it undergoes a rescue with constant rate $r_r$, switching back to the growing state. 

Every time the plus end reaches a transverse microtubule - i.e. when its position is $x = nd$ with $n \in \mathbb{N}$, it creates a crossover. This crossover can be resolved in two distinct ways: either it is removed by the shrinkage of the microtubule due to its dynamic instability, or it survives long enough to lead to a severing event.  Whether or not the severing event occurs is determined not only by the dynamic instability of the longitudinal microtubule, but also by an intrinsic severing waiting time distribution at the crossover that can be, in principle, arbitrary. Here, however, we choose a distribution that best fits the biological experimental data about the action of \emph{katanin} at crossovers \cite{Lindeboom2018CLASPReorientation.}. Moreover, we need the distribution to account the fact that \emph{katanin} requires a certain amount of time to localize at crossovers before being able to sever microtubules. For this reason we choose the severing waiting time distribution to be
\begin{equation} \label{gamma_sever-wait_distr}
W_{k \theta} \left( t \right) = \frac{t^{k-1}}{\theta^k \Gamma\left(k\right)} e^{-\frac{t}{\theta}},
\end{equation}
i.e. Gamma probability density function \cite{Papoulis2002ProbabilityProcesses}, where
\begin{equation}
\Gamma \left( k \right) = \int_0^\infty ds \, s^{k-1} e^{-s},
\end{equation}
is the Euler gamma function, $k$ is the shape, and $\theta$ is the scale parameter of the distribution.

When the severing event occurs, the former long microtubule is split in two shorter microtubules, and both of them keep undergoing dynamic instability and can create new crossovers and being severed again, in order to amplify the number of longitudinal microtubules. The newly-created plus end of the lagging microtubule either is stabilized and it enters the growing state with probability $p^+$, or it enters the shrinking state with probability $1-p^+$. The newly-created minus end of the leading microtubule is now positioned at the severing point in a stable state, whilst no changes are applied to its plus end, see Figure \ref{fig1_model}.

\subsection{Microtubule behaviour in the interstitial strip}

After the creation of a crossover and before the creation of a second one, the dynamics of the plus end of a microtubule is described by the Dogterom-Leibler model for microtubules undergoing dynamic instability \cite{Dogterom1993PhysicalStructures}. Notice that the dynamics of the plus end is not influenced by eventual severing events. Therefore, as long as the plus end is at $x \in \left(nd, (n+1)d \right)$, we can study the property of the correspondent microtubule undergoing dynamic instability in a strip of width $d$ as if its length is $l=x-nd$.

%If $m^\sigma \left( t, l \right)$ is the probability distribution for a microtubule with length $l$ and in state $\sigma = \pm$ for respectively growing and shrinking microtubules at time $t$, then it satisfies
%\begin{equation}     \label{dynamic-eqs_+}
%\frac{\partial}{\partial t} m^+ \left( t, l \right) = - v^+ \frac{\partial}{\partial l} m^+ \left( t, l \right) - r_c m^+ \left( t, l \right) + r_r m^- \left( t, l \right),
%\end{equation}
%\begin{equation}     \label{dynamic-eqs_-}
%\frac{\partial}{\partial t} m^- \left( t, l \right) = v^- \frac{\partial}{\partial l} m^- \left( t, l \right) - r_r m^- \left( t, l \right) + r_c m^+ \left( t, l \right).
%\end{equation}
%In the non-confined case, this set of equations has two possible solutions: in the bounded-growth regime, defined by the relation $\overline{l} > 0$, with
%\begin{equation}     \label{typical-length}
%\overline{l} = \left( \frac{r_c}{v^+} - \frac{r_r}{v^-} \right)^{-1},
%\end{equation}
%the steady-state solution is reached, and both densities are exponential decays $m^\sigma \left( l \right) \propto e^{-l/\overline{l}}$, whilst in the unbounded-growth regime the average length of microtubules grows linearly in time, with the densities that are well-approximated by Gaussian-like functions \cite{Dogterom1993PhysicalStructures}.

In the non-confined-in-a-strip case, the model has two possible solutions for the probability distribution of microtubule length: in the bounded-growth regime, defined by the relation $\overline{l} > 0$, with
\begin{equation}     \label{typical-length}
\overline{l} = \left( \frac{r_c}{v^+} - \frac{r_r}{v^-} \right)^{-1},
\end{equation}
the steady-state solution is reached, and the length distribution is an exponential decay proportional to $e^{-l/\overline{l}}$, whilst in the unbounded-growth regime the average length of microtubules grows linearly in time, with the length distribution that is well-approximated by a Gaussian-like function \cite{Dogterom1993PhysicalStructures}.

When microtubule dynamics is confined in a strip of a finite width, however, both the bounded and the unbounded-growth regimes lead to a steady-state solution for the length distribution that is proportional to $e^{-l/\overline{l}}$. Notice that, in the unbounded-growth regime case, $\overline{l}$ is no-longer positive, and hence the distribution is exponentially increasing \cite{Govindan2004SteadyGeometry}. General features regarding the lifetime distribution and the splitting probabilities of microtubules in the Dogterom-Leibler model can be found in the Appendix A.

Given our specific interest in studying the properties of the system in both the bounded and the unbounded-growth regime, we have chosen two sets of dynamic parameters: for the bounded-growth case parameters are chosen accordingly to previous observations \cite{Vos2004MicrotubulesTranslocation}, while for the unbounded-growth case, both dynamic parameters and grid parameters are those that have been directly measured for the WT case by previous experimental works \cite{Lindeboom2018CLASPReorientation.}, see Table \ref{table1-parameters}.

\begin{table}
  \begin{center}
  \begin{tabular}{c c c c c}
    \hline
    \hline
    \multirow{2}{*}{Parameter} & \multirow{2}{*}{Description} & Numerical value & Numerical value & \multirow{2}{*}{Units} \\
    & & (bounded-growth) & (unbounded-growth) & \\
    \hline
    $v^+$ & Growth speed & $0.1$ & $0.103$ & $\mu\mbox{m} \, \mbox{s}^{-1}$ \\
    $v^-$ & Shrinkage speed & $0.25$ & $0.225$ & $\mu\mbox{m} \, \mbox{s}^{-1}$ \\
    $r_c$ & Catastrophe rate & $0.02$ & $0.0058$ & $\mbox{s}^{-1}$ \\
    $r_r$ & Rescue rate & $0.02$ & $0.026$ & $\mbox{s}^{-1}$ \\
    $p^+$ & Probability of rescue-after-severing & Tuned & Tuned & - \\
    $d$ & Spacing between neighbors & $1.5$ & $1.5$ & $\mu\mbox{m}$ \\
    $\theta$ & Scale parameter of Gamma distribution & $8.5$ & $8.5$ & s \\
    $k$ & Shape parameter of Gamma distribution & $7$ & $7$ & - \\
    \hline
    \hline
  \end{tabular}
  \caption{Model parameters.}
  \label{table1-parameters}
  \end{center}
\end{table}

\section{Results}
\subsection{Amplification in the unbounded-growth regime}

The model introduced in the last section had partially been computationally studied for microtubules in the unbounded-growth regime, and it shows that the factor that influences the most the speed of the amplification of longitudinal microtubules is the probability of rescue-after-severing $p^+$ rather than the intrinsic rescue rate $r_r$ of microtubules \cite{Lindeboom2018CLASPReorientation.}. 

Here, we want to perform an in-depth study of the response to the system to the change of $p^+$. We will show that, even though $p^+$ is crucial for the speed of amplification, in the unbounded-growth regime it is not required for the occurrence of it.

Our simulations consist of $N = 10^5$ trials in which a single longitudinal microtubule undergoes dynamic instability in the whole grid of transverse microtubules. For every trial we keep track of the fate of the initial microtubule and its offspring until either no more microtubules are present - i.e. they all have shrunk to length zero, and we call this possible output extinction, or, for every trial that did not result in an extinction, the number of microtubules is exponentially increasing, and we call this second possible output amplification.

\begin{figure}[htbp]   
  \centering
  \includegraphics[scale=0.27]{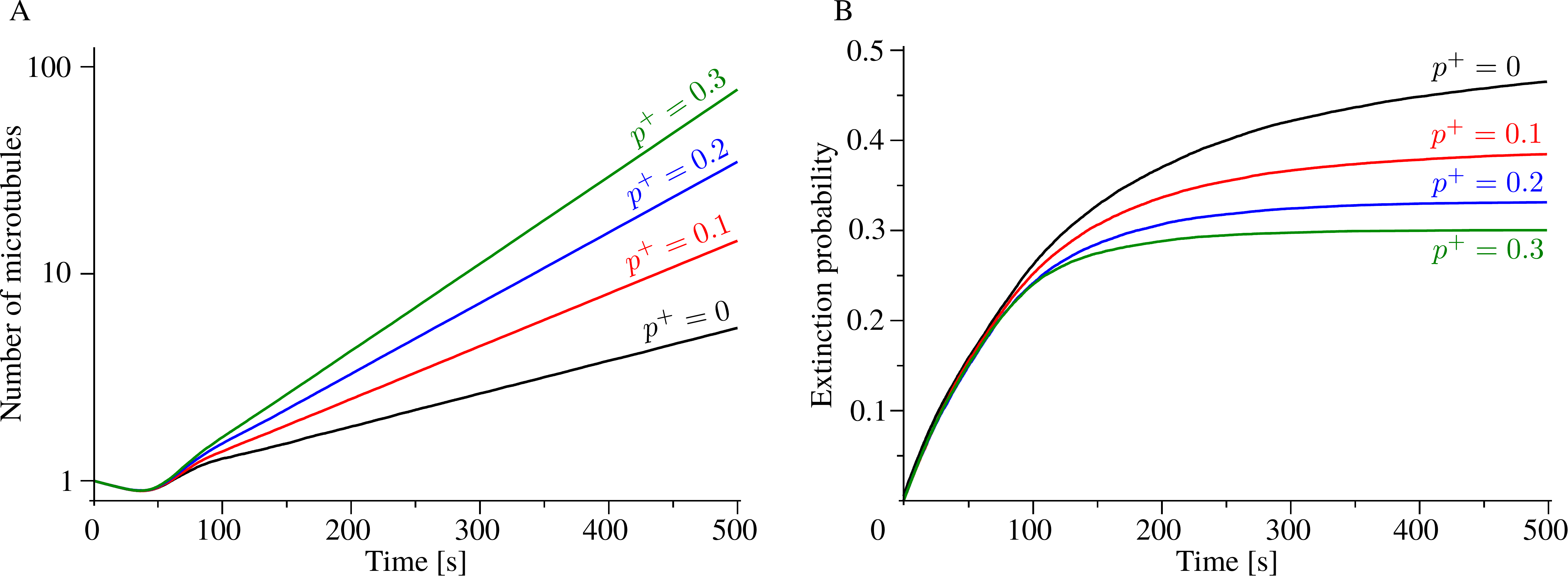}
  \caption{(A) Time evolution of the number of longitudinal microtubules for four different values of $p^+$. They all exhibit amplification. (B) Extinction probability as a function of time. It represents the fraction of trials in which, after a certain amount of time, all microtubules have completely depolymerized. }  \label{fig2A_amplification-extinction_unbounded}
\end{figure}

Fig. \ref{fig2A_amplification-extinction_unbounded}A shows that for our choice of dynamic parameters, the speed of amplification increases with $p^+$. Furthermore, we notice from Fig. \ref{fig2A_amplification-extinction_unbounded}B that greater values of $p^+$ correspond to lower extinction probabilities, suggesting that a good rescue-after-severing entails a double effect: not only it increases the speed of amplification, but also makes the amplification occur more likely.

%It is also interesting to notice that, as $p^+$ raises, the extinction probability reaches a plateau, i.e. the number of microtubules generated by the first one and its descendants is great enough to prevent the system from an extinction. We call this plateau \textit{ultimate extinction probability}. The ultimate extinction probability can be used as an order parameter to estimate how many microtubules are required to be sure that the amplification occurs.

The interesting result that in the unbounded-growth regime even the $p^+=0$ case leads to an overall amplification can be explained by an intrinsic property of the regime itself. Indeed, although every severing event shortens the length of the severed microtubule, its plus end is not affected by such an event. Consequently, the dynamic properties of the leading microtubule are not changed by the severing, and so it applies to the microtubule lifetime as well. Since, on average, the length of microtubules in the unbounded-growth regime grows as
\begin{equation}
J = \frac{ r_r v^+ - r_c v^-}{r_r + r_c} t,
\end{equation}
it follows that the average lifetime of microtubules is infinite \cite{Dogterom1993PhysicalStructures}, and therefore there is no upper bound for the number of severing events that a microtubule can undergo.

\subsection{Amplification in the bounded-growth regime}

In this section, we address the question whether or not the amplification occurs regardless of $p^+$ in the bounded-growth regime as well as in the unbounded-growth case. To do so we perform computer simulations for microtubules in the bounded-growth regime (see Table \ref{table1-parameters}) to show that $p^+$ needs to be greater that a certain critical value $p^+_{crit}$ in order have amplification. Moreover, using a combination of computer simulations and analytical calculations we identify such a critical value as a function of the other model parameters.

\subsubsection{Critical point in simulations}

\begin{figure}[htbp]   
  \centering
  \includegraphics[scale=0.27]{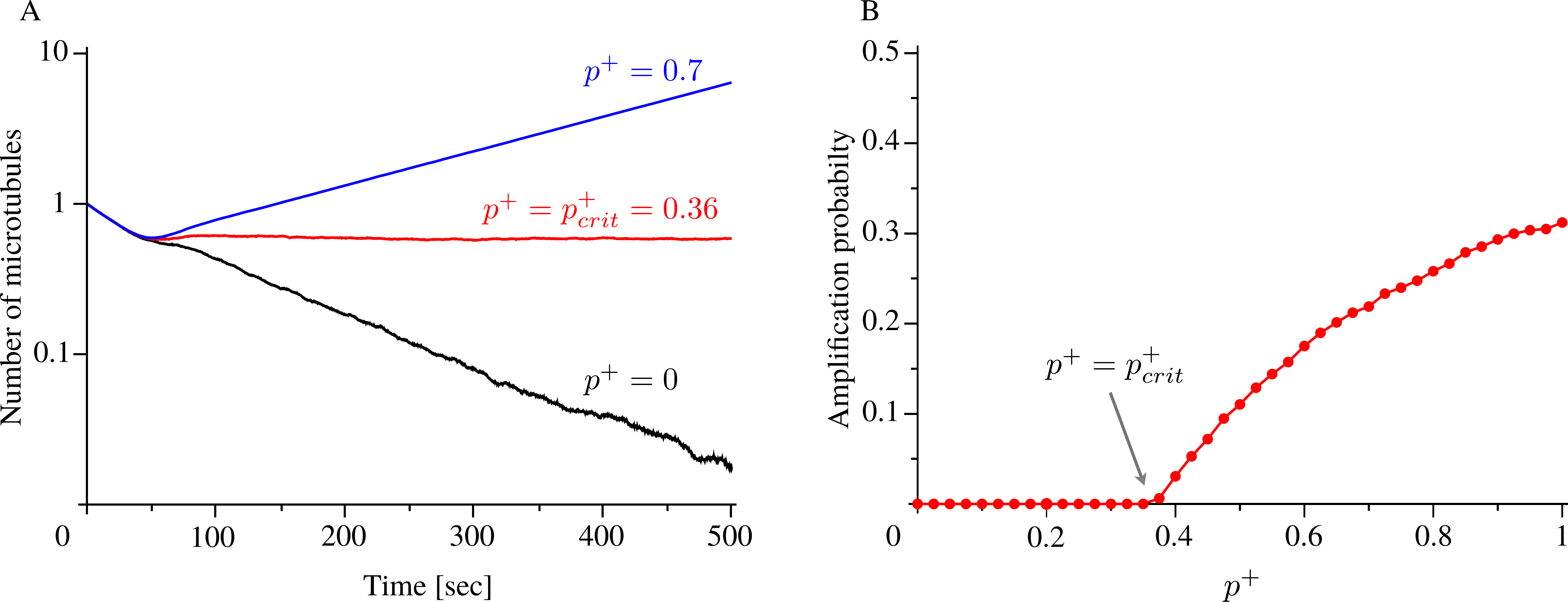}
  \caption{(A) Time evolution of the number of longitudinal microtubules for three different values of $p^+$. One leads to amplification (blue line), one to extinction (black line), and one corresponds to the critical behaviour (red line). (B) Amplification probability as a function of $p^+$. Amplification probability is non-zero for $p^+$ larger than $p^+_{crit} \simeq 0.36$. }  \label{fig2_amplification-criticality}
\end{figure}

By tuning the probability of rescue-after-severing $p^+$ from $0$ to $1$, we observe two different behaviours, see Figure \ref{fig2_amplification-criticality}A: for lower values of $p^+$ the average number of microtubules exponentially decays in time (extinction), whilst for higher values of $p^+$ the number of microtubules exponentially increases (amplification). It follows that there exists a critical threshold for $p^+$ above which the average output is amplification. For our choice of model parameters, the computationally measured critical value is $p^+_{crit} \simeq 0.36$. Therefore, if we define the \textit{amplification probability} as the fraction of trials the output of which is amplification, we observe that below the critical threshold the amplification probability is zero, whilst it is greater than zero otherwise, see Figure \ref{fig2_amplification-criticality}B. %Notice that our definition of the amplification probability implies that it is the complementary of the ultimate extinction probability.

\subsubsection{Calculation of the critical point}

When microtubule is created through a severing event, either it shrinks to length zero and dies, or it is severed a sufficient number of times to create an offspring of new lagging microtubules. If the size of such an offspring is, on average, greater than one, the output is amplification. In other words, if $M$ is the number of severing events that a newly-created microtubule undergoes, then amplification occurs if, on average,
\begin{equation}    \label{amplification-condition_1}
M > 1.
\end{equation}

%The amplification mechanism can be seen as the product of a chain reaction. Indeed, every severing event creates new lagging microtubules that act as intermediates of the reaction during the propagation step (Laidler, Chemical Kinetics, 1987). The newly-created lagging microtubule can either shrink to length zero and die, or create a offspring of new lagging microtubules. If the size of the offspring is, on average, greater than one, the output is amplification. In other words, suppose that a newly-created microtubule is severed $M$ times (after the first severing event it becomes a leading microtubule). Then amplification occurs if, on average,
%\begin{equation}    \label{amplification-condition_1}
%M > 1.
%\end{equation}

To fix the ideas, suppose that a microtubule is created by severing with initial length $x=d$. Then, with probability $p^+$ it is initially created in the growing state, and consequently with probability $1-p^+$ in the shrinking state. It follows that the size of the offspring of the mother microtubule can be written as $M = p^+ M^+ + \left( 1- p^+ \right) M^-$, where $M^\sigma$ is the size of the offspring of a  microtubule created in the state $\sigma$. However, since a shrinking microtubule with plus end at $x < d$ cannot be severed, $M^-$ equals $M^+$ times the probability that the shrinking microtubule recovers the length at birth $d$, i.e. $M^- = R^-_{d} \left( d \right) M^+$, where $R^-_{d} \left( d \right)$ is the splitting probability defined in the Appendix A. Hence, the condition expressed in Eq. (\ref{amplification-condition_1}) can be rewritten as
\begin{equation}    \label{amplification-condition_2}
M = \left[ p^+ + \left( 1 - p^+ \right) R^-_{d} \left( d \right) \right] M^+ > 1.
\end{equation}
By solving the equality related to Eq. (\ref{amplification-condition_2}) we can find the critical value of $p^+ = p^+_{crit}$ above which amplification takes place
\begin{equation}     \label{p+_crit}
p^+_{crit} = \frac{1 - R^-_{d} \left( d \right) M^+}{ \left( 1 - R^-_{d} \left( d \right) \right) M^+ }.
\end{equation}

From this equation, we can identify the two extreme scenarios in which amplification never or always occurs, regardless of $p^+$. In the first case, we state that amplification never occurs if $p^+_{crit} > 1$, meaning that the maximum value that $p^+$ can reach is not enough to lead to amplification. We can show that condition $p^+_{crit} > 1$ is equivalent to condition $M^+ < 1$, i.e. amplification is impossible if the average size of the offspring of mother microtubules born in growing state is smaller than $1$. In the second case, we state that amplification always occurs if $p^+_{crit} < 0$, meaning that even without rescue-after-severing, dynamic parameters of the model are such that amplification is still possible. This condition is equivalent to $M^+ R^-_{d} \left( d \right) > 1$, or $M^- >1$, i.e. amplification occurs every time the average size of the offspring of mother microtubules born in shrinking state is greater than $1$.

It is important to underline that, in our discussion, we assumed that all microtubules were born with initial length $d$. This choice implies that all severing events occur at the first crossover. 
%We refer to this approximation as \textit{first-crossover} approximation. 
However, given the stochastic nature of the system and of the severing waiting time probability of Eq. (\ref{gamma_sever-wait_distr}), it is possible that a severing event occurs further in the grid than at the first crossover of a microtubule. In other words, the initial length of a newly-created microtubule can be $x = n d$, with $n>1$.

In this case, we need to add into the count of the size of offspring of a mother microtubule all cases in which a microtubule that is born with initial length $n d$, $n>1$, it is also severed at $(n-1) d$, $(n-2) d$, $\dots$ . We consider the microtubules created by this mechanism as direct daughter microtubules of the mother microtubule we are measuring the size of the offspring of. To that end, we first define $m_i$ via $M^+ = \frac{1}{N} \sum\limits_{i=1}^N m_i$ as the number of microtubules generated by the mother microtubule labeled by $i$, and $N$ is the number of microtubules we keep track of the fate. Then, we denote the number of severing events the microtubule $i$ undergoes with $s_i$, and the position of the crossover at which the first severing takes place with $c_{j_i}$, with the rule: $c_{j_i} = n-1$ if the severing occurred at $nd$. Since after a severing event at $nd$ the crossovers at $kd$, $k<n$, can be removed by either shrinkage or severing, we define $b_{c_{j_i}}$ as the number of crossovers that are resolved by shrinkage. Therefore
\begin{equation}       \label{m_i}
m_i = s_i + \sum_{j_i=1}^{s_i} \left[ c_{j_i} - b_{c_{j_i}} \right].
\end{equation}
Notice that $b_{c_{j_i}}$ depends on $p^+$, as it depends on the behaviour of the plus end after severing of the newly-created microtubule. Consequently, the r.h.s. of Eq. (\ref{p+_crit}) exhibits a dependency on $p^+$. Hence, we need to find the exact dependency on $p^+$ of $b_{c_{j_i}}$. To avoid the problem, in first approximation we set $b_{c_{j_i}} = 0$ for every $c_{j_i}$. In this way, we can computationally measure $m_i^{(1)} \equiv s_i + \sum\limits_{j_i=1}^{s_i} c_{j_i}$, and we use Eq. (\ref{p+_crit}) to give a first estimate of the critical probability $p^+_{crit,(1)}$, see Table \ref{table2-crit-p+}. We refer to this approximation as \textit{one-crossover} approximation. From the table we notice that, although this approximation gives a reasonable prediction for the critical probability of rescue-after-severing, it systematically underestimates it.

Analytically, one can calculate $b_{c_{j_i}}$ only under the condition that a severing event at $nd$ always implies the resolution of the crossovers at $d$, $2 d$, $\dots$, $\left( n - 2 \right) d$ through a severing event, whilst the crossover at $\left( n - 1 \right) d$ can be resolved by either severing or shrinkage. 
%We refer to this approximation as \textit{two-crossovers} approximation. 
Therefore, if we denote the probability of resolving this crossover with a shrinkage as $p_{cr} \left( p^+ \right)$, we have
\begin{equation}     \label{b_c_j_i}
b_{c_{j_i}} = \left( 1 - \delta_{c_{j_i},0} \right) p_{cr} \left( p^+ \right).
\end{equation}
The Kronecker function $\delta_{c_{j_i},0}$ of Eq. (\ref{b_c_j_i}) accounts the fact that, if the severing happens at $d$ (i.e. $c_{j_i} = 0$), no other crossovers are removed by either severing or shrinkage. If we plug Eq. (\ref{b_c_j_i}) into Eq. (\ref{m_i}), we can now calculate an approximate expression for $M^+$, i.e.
\begin{equation} \begin{split}    \label{M+1}
M^+_{(2)} & = \frac{1}{N} \sum_{i=1}^N m_i^{(2)} = \frac{1}{N} \sum_{i=1}^N \Big[ s_i + \sum_{j_i=1}^{s_i} c_{j_i} - p_{cr} \left( p^+ \right)  \sum_{j_i=1}^{s_i} \left( 1 - \delta_{c_i,0} \right) \Big].
%& = M^+_{(1)} - p_{cr} \left( p^+ \right) S \frac{1}{N} \sum_{i=1}^N  \left\langle 1 - \delta_{c_i,0} \right\rangle,
\end{split} \end{equation}
The detailed derivation of Eq. (\ref{M+1}) can be found in the Appendix B. Finally, if we define $M^+_{(1)} = \frac{1}{N} \sum\limits_{i=1}^N \left[ s_i + \sum\limits_{j_i=1}^{s_i} c_{j_i} \right]$, $S = \frac{1}{N} \sum\limits_{i=1}^N s_i$, and $\left\langle 1 - \delta_{c_i,0} \right\rangle = \frac{1}{s_i} \sum\limits_{j_i=1}^{s_i} \left( 1 - \delta_{c_{j_i},0} \right)$, (see Appendix B), we can rewrite the condition (\ref{amplification-condition_2}) for microtubule amplification as
\begin{equation}    \label{amplification-condition_final}
M = \Big[ p^+ + \left( 1 - p^+ \right) R^-_{d} \left( d \right) \Big] \left[ M^+_{(1)} - p_{cr} \left( p^+ \right) S \frac{1}{N} \sum_{i=1}^N  \left\langle 1 - \delta_{c_i,0} \right\rangle \right] > 1.
\end{equation}
The resolution of the equation associated to this inequality provides the critical threshold for the probability of rescue-after-severing $p^+_{crit,(2)}$ in order to have amplification. We refer to this approximation as \textit{two-crossovers} approximation. The expression (\ref{amplification-condition_final}) contains two quantities, $M_{(1)}^+$ and $S$, that cannot be analytically calculated but can be easily measured with computer simulations. On the other hand, in the following sections we are going to analytically calculate the terms $p_{cr} \left( p^+ \right)$ and $\frac{1}{N}\sum\limits_{i=1}^N  \left\langle 1 - \delta_{c_i,0} \right\rangle$. In this way, we will make a better prediction off the critical probability of rescue-after-severing in order to have amplification.

\subsection{Analytical approach}

In this section we are going to calculate the critical probability of rescue-after-severing. 
%Our starting point is the inequality (\ref{amplification-condition_final}). From the equation associated to that condition, it follows that we need to analytically calculate the probabilities $p_{cr} \left( p^+ \right)$ and $\frac{1}{N}\sum\limits_{i=1}^N  \left\langle 1 - \delta_{c_i,0} \right\rangle$, i.e. the probability that a crossover is removed by the shrinkage of a microtubule induced by a severing event at a following crossover, and the probability that a microtubule is severed at a crossover different than the first one, respectively. 
To that end, we first need to calculate $p_{cr} \left( p^+ \right)$ and $\frac{1}{N}\sum\limits_{i=1}^N  \left\langle 1 - \delta_{c_i,0} \right\rangle$. In order to do so, and because of the complexity of the model, we make the approximation that the entire grid of transverse microtubules is replaced by just two transverse microtubules. This reduces the total number of possible crossovers to two. Therefore, we first calculate the FPTD for a longitudinal microtubule to create a crossover with a transverse, as we will need it for the formulation of our \textit{two-crossovers approximation}. Then, we give some analytical results of the one-crossover approximation already introduced in the previous section. Finally, we present the two-crossovers approximation and we show that we can use it to calculate the critical probability of rescue-after-severing with a good degree of accuracy.

\subsubsection{The first passage time distribution}

The creation of new crossovers for a microtubule undergoing dynamic instability is intimately linked to a FPTD problem for the same microtubule to reach a target. Here, we face this problem by making use of an approach where we consider all possible legal paths to reach the target, given the knowledge of the time needed to reach it.

\begin{figure}[htbp]   
  \centering
  \includegraphics[scale=0.27]{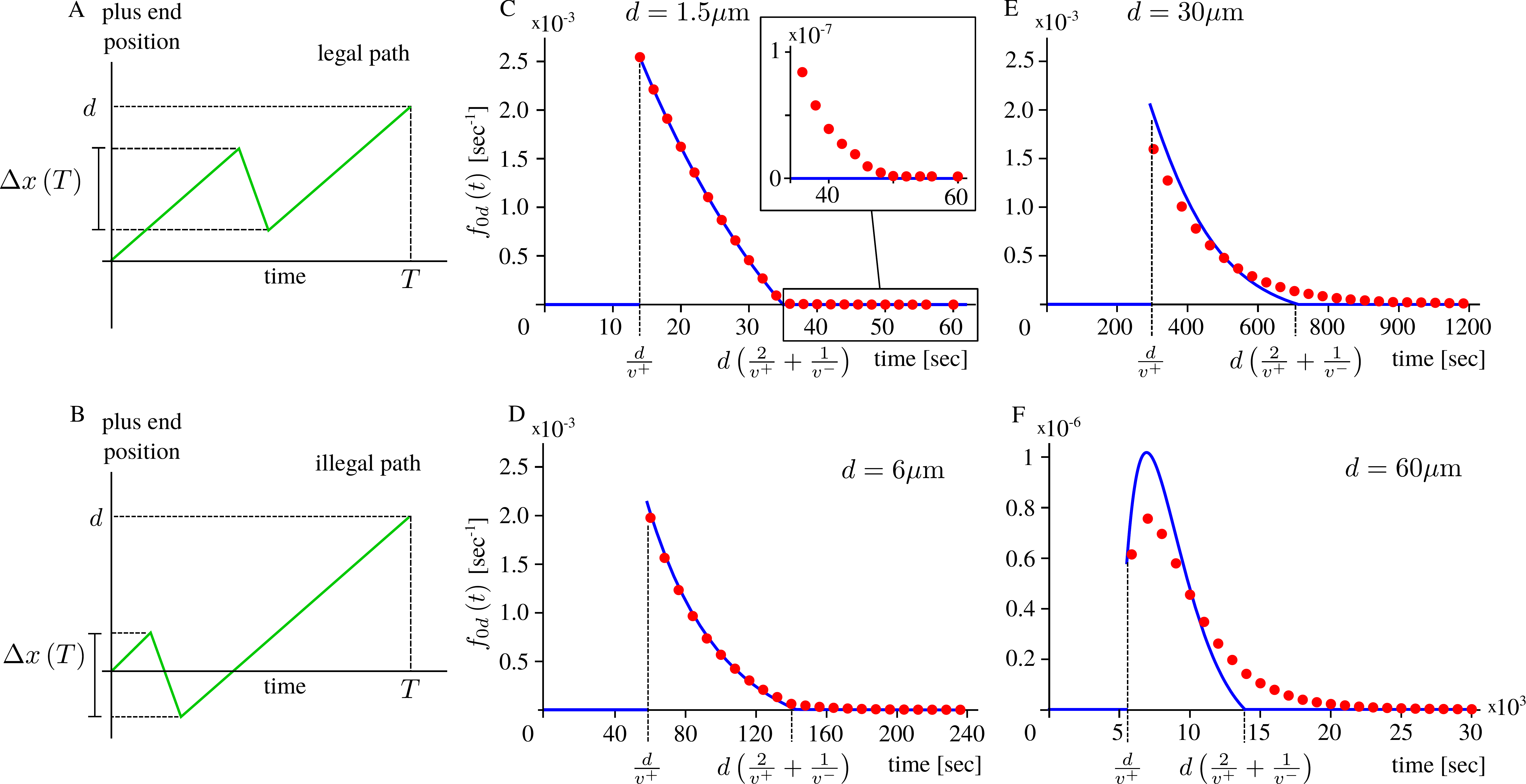}
  \caption{(A) Legal and (B) illegal path for a microtubule to reach the target at a distance $d$ in a first passage time $T$. Only one catastrophe and one rescue are allowed. (C-F) Comparison between simulations (red dots) and theory (blue line) for the non-direct part $f_d \left( t \right)$ of the first passage time distribution. (B, C) Our theory nicely fits simulations for relatively close targets ($d = 1.5 \, \mu\mbox{m}$, $d = 6 \, \mu\mbox{m}$), (D, E) while it is not very good for more distant targets ($d = 30 \, \mu\mbox{m}$, $d = 60 \, \mu\mbox{m}$).}  \label{fig4_completeFPTD}
\end{figure}

The first passage time problem for a microtubule to reach length $x_1$ starting from $x_0$ in the absence of severing can be seen as a reverse lifetime problem, in the sense that in place of studying the reaching of the target at $x_1$ we study the survival of the microtubule until it arrives at $x_1$, as if it is shrinking from $x_0$ to $x_1$. In this way, the growing speed of the microtubule acts as its shrinking speed, its catastrophe rate as the rescue rate, and viceversa. However, with this approach we assume that a microtubule ``shrinking'' from $x_0$ can undergo a ``rescue'' and grow beyond the initial position $x_0$. This means that the microtubule assumes negative length. To avoid this, we need to take into account only the paths from $x_0$ to $x_1$ that never shrink below $x_0$. Hence, if $L_\sigma \left( t | x_1 -x_0 \right)$ is the lifetime distribution for microtubules with initial length $x_1 - x_0$ and initial state $\sigma$ (see Appendix A), we define 
\begin{equation}         \label{lifetime-distribution_-_reverse}
L_\sigma^{target} \left( t, x_1 - x_0 \right) = \big. L_\sigma \left( t | x_1 - x_0 \right) \Big|_{\tiny{\begin{matrix} v^\pm \to v^\mp \\ r_c \leftrightarrow r_r \end{matrix}}},
\end{equation}
as the FPTD to reach the target, including the possibility of assuming negative length. Therefore, this function must be re-scaled by the number of \textit{legal paths} $\Gamma_{x_0 \to x_1} \left( t \right)$ that reach the target $x_1$ at time $t$, without ever shrinking back to $x<x_0$, calculated over all possible paths that arrive at $x_1$ at time $t$, see Figure \ref{fig4_completeFPTD}AB.

For our purpose, the target to reach is a transverse microtubule for the creation of a new crossover, the position of which is at distance $d$ from the starting point, i.e. the previous transverse microtubule. Typically, for the range of values of Table \ref{table1-parameters}, every plus end that impinges on a transverse microtubule starting from the previous one either it does it without undergoing any catastrophe, or it undergoes one catastrophe and a subsequent rescue. For the dynamic parameters we are considering, the occurrence of multiple catastrophe-rescue events is very unlikely. Therefore we assume that all paths are either direct - no catastrophes, or indirect - one catastrophe and one rescue.

Given the constant growing and shrinking speeds, the amount of time that a microtubule needs to reach the target at $d$ is given by the time needed to reach it in absence of any catastrophe, added to the time spent from a catastrophe to the moment when the original length before the catastrophe is restored. Mathematically, if $T_d$ is the first-passage time and $\Delta x \left( T_d \right)$ is the distance walked by the plus end from the catastrophe to the subsequent rescue, the equation
\begin{equation}         \label{FPT}
T_d = \frac{d}{v^+} + \Delta x \left( T_d \right) \left( \frac{1}{v^+} + \frac{1}{v^+} \right),
\end{equation}
holds. From Eq. (\ref{FPT}) we can find the expression for $\Delta x \left( T \right) = \frac{v^+ v^-}{v^+ + v^-} \left( T - \frac{d}{v^+}\right)$.

Since catastrophes are modelled as Poisson events, if a catastrophe occurs, the probability that it occurs does not depend on the distance from the target. Therefore, the fraction of legal paths can be written as $\Gamma_{0 \to d} \left( T \right) = 1 - \frac{\Delta x \left( T \right)}{d}$, and, finally, the FPTD as
\begin{equation}       \label{FPTD}
F_{0d} \left( t \right) = L_\sigma^{target} \left( t, d \right) \Gamma_{0 \to d} \left( t \right) \Theta \left[ d \left( \frac{2}{v^+} + \frac{1}{v^-} \right) - t \right],
\end{equation}
where the Heaviside theta is imposed to allow at most one catastrophe-rescue event. %Indeed $t = d \left( \frac{2}{v^+} + \frac{1}{v^-} \right)$ is the amount of time that a microtubule needs to almost reach the target but without touching it, i.e. $\frac{d}{v^+}$, then undergo a catastrophe and almost shrink to length $0$, i.e. $\frac{d}{v^-}$, and finally undergo a rescue and grow until the target is reached, $\frac{d}{v^+}$.

In order to separate direct paths from indirect paths, it is convenient to split $F_{0d} \left( t \right)$ in two parts, and rewrite it as
\begin{equation}       \label{FPTD-split}
F_{0d} \left( t \right) = \delta \left( t - \frac{d}{v^+} \right) \, e^{-r_c t} + f_{0d} \left( t \right),
\end{equation}
where the term multiplied by the delta function accounts direct paths, while $f_{0d} \left( t \right)$ accounts indirect. From Appendix A that microtubules reach the target with probability $R_d^+ \left( x \right)$. Therefore $F_{0d} \left( t \right)$ is normalized to $R_d^+ \left( x \right)$, and as a consequence, the relation
\begin{equation}
\int_0^\infty dt \, f_{0d} \left( t \right) = R_d^+ \left( 0 \right) - e^{-\frac{r_c d}{v^+}}
\end{equation}
holds.

We run $N = 10^6$ simulations of microtubules undergoing dynamic instability in a strip of width $d$, and we create the histogram of the arrival times for the microtubules that reach the target with an indirect path. Figure \ref{fig4_completeFPTD}CD shows that the approximation of only one catastrophe-rescue event is a good approximation when the target is relatively close compared to the dynamic parameters of the microtubules, while it apparently fails when the target is more distant, see Figure \ref{fig4_completeFPTD}EF. However, it is convenient to point out that, for $d \gg \overline{l}$, we observe a very few arrivals at the target, since from Eq. (\ref{splitting-xtod+}) we notice that the arrival probability $R_d^+ \left( 0 \right)$ decays as $e^{-d/\overline{l}}$. On the other hand, in the unbounded-growth regime, since a fraction $1-\frac{r_c v^-}{r_r v^+}$ of the microtubules always arrives at the target, for distant targets the approximation of one catastrophe-rescue event is no longer accurate.

\subsubsection{One-crossover theory}

Na\"ively, one can think that once a crossover is created, the probability $p^{(1)}_{sev}$ of resolving it with a severing event is given by the competition between two independent events: microtubule lifetime, expressed by the random variable $T_+ \left( x \right)$ with density function given by Eq. (\ref{lifetime-distribution_+}), and severing waiting time at the crossover, with random variable $\tau_d$ and density function defined in Eq. (\ref{gamma_sever-wait_distr}). Then, if we define the random variable $t = \tau_d - T_0$, we can calculate its probability density function by using the relation $P_{z} \left( z = x + y \right) = \int_{-\infty}^{+\infty} dz' \, P_x \left( z - z' \right) P_y \left( z' \right),$
%\begin{equation}       \label{sum-RV_pdf}
%P_{z} \left( z = x + y \right) = \int_{-\infty}^{+\infty} dz' \, P_x \left( z - z' \right) P_y \left( z' \right),
%\end{equation}
where $P_i$ is the probability density function of the random variable $i = x,y,z$, and $x$ and $y$ are independent random variables. In our case, the probability density function is
\begin{equation}      \label{taud-T0_pdf}
P_{\tau_d - T_+ \left( 0 \right)} \left( t \right) = \int_{-\infty}^{+\infty} dt' \, W_{k,\theta} \left( t + t' \right) L_+ \left( t' | 0 \right).
\end{equation}
Hence, the probability that the event ``severing'' happens before the event ``return'' is $\mathbb{P}\left[ t < 0 \right] = \int_{-\infty}^{0} dt \, P_{\tau_d - T_+ \left( 0 \right)} \left( t \right)$. This probability is not yet the probability of resolving a crossover with a severing event: indeed, microtubules in the unbounded-growth regime have a finite probability of growing indefinitely. For those, the lifetime $T_+ \left( 0 \right) \to \infty$ is infinite. Therefore, the probability of resolving a crossover with a severing event is
\begin{equation}          \label{psevONE}
p^{(1)}_{sev}  = S_+ \left( \infty | 0 \right) + \left[ 1 - S_+ \left( \infty | 0 \right) \right] \int_{-\infty}^{0} dt \, P_{\tau_0 - T_+ \left( 0 \right)} \left( t \right),
\end{equation}
where $S_+ \left( \infty | 0 \right)$ is the ultimate survival probability defined in the Appendix A. As a consequence the probability of resolving a crossover with a shrinkage is
\begin{equation}          \label{pshrinkONE}
p^{(1)}_{shrink}  = 1 - p^{(1)}_{sev}.
\end{equation}

However, with this approach we neglect the number of crossovers removed by shrinkage after the severing at a second crossover, and, therefore, the dependency on $p^+$. In other words, the one-crossover theory does not take into account that some microtubules that would have been severed at $d$ are not anymore severed there because an eventual severing at $nd$, $n>1$, can in principle shorten their lifetimes, and make them shrink below $d$, resulting in the resolution of the crossover by a shrinkage induced by the severing at $nd$, see Figure \ref{fig5_sevVSshrink_V3}.

\subsubsection{Two-crossovers theory}

\begin{figure}[htbp]   
  \centering
  \includegraphics[scale=0.27]{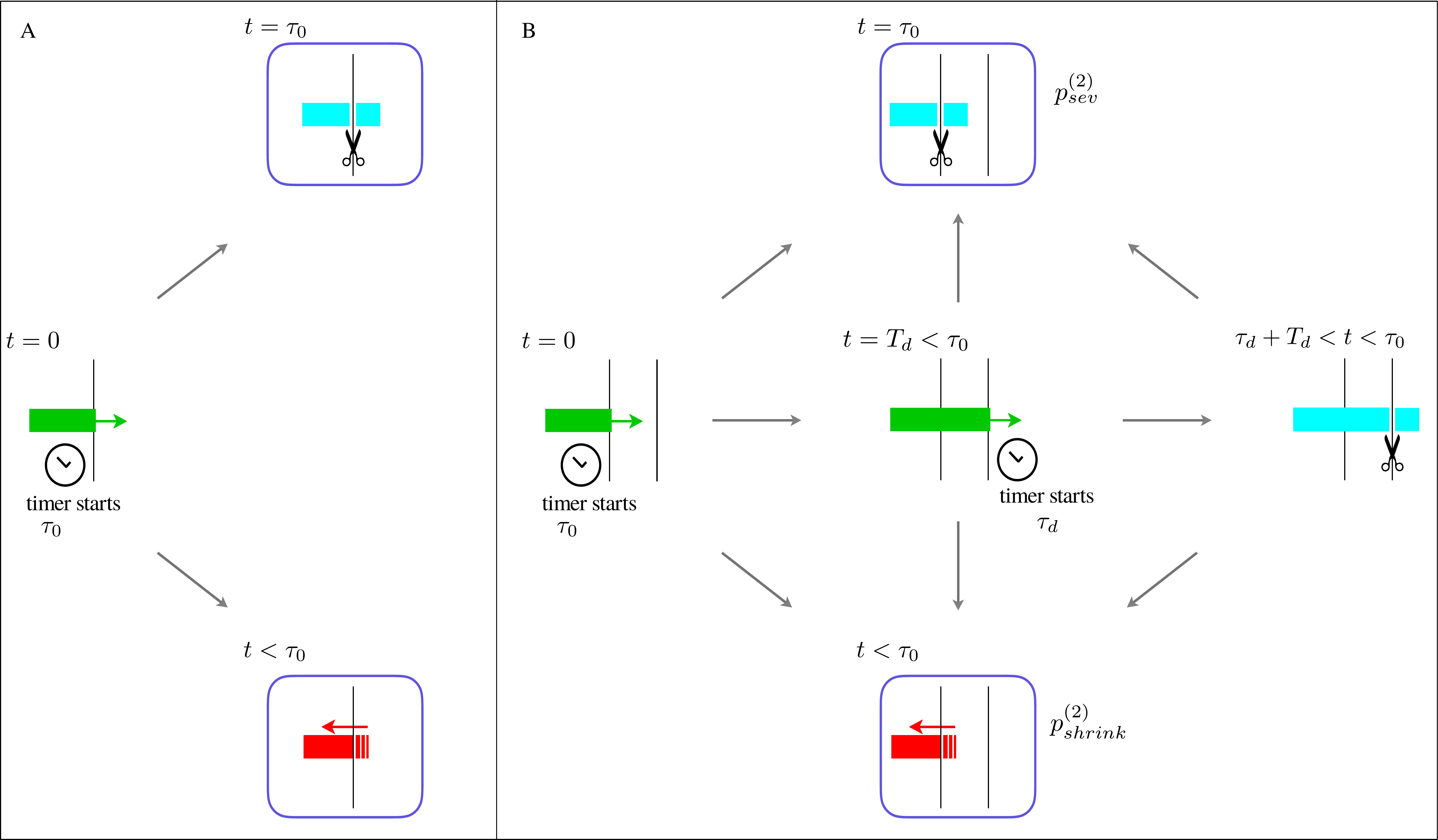}
  \caption{Schematic of the full one-crossover theory (A). The newly-created crossover can be resolved either by the shrinkage of the plus end below the crossover itself (lower blue square), or by the severing at crossover (upper blue square). Schematic of the full two-crossovers theory (B). The first crossover created can be resolved either by the shrinkage of the longitudinal microtubule with probability $p_{shrink}^{(2)}$ (sum of all paths that bring to the lower blue square), or by the severing at crossover with probability $p_{sev}^{(2)}$ (sum of all paths that bring to the upper blue square). Whether the severing at the first crossover occurs or not also depends on what happens at the second crossover: a severing event at the second crossover alters the dynamic instability of the lagging microtubule, and hence its probability of shrinking before being severed at the first crossover.}  \label{fig5_sevVSshrink_V3}
\end{figure}

In order to take into account the influence of a crossover on the resolution of the previous one, we calculate the probability of resolving a crossover with a severing event in a scenario in which we have two transverse microtubules, at position $d$ and $2d$ respectively. We make the further approximation that a microtubule cannot be severed two times at the same crossover. We denote the probability of having a severing event at $d$ with $p^{(2)}_{sev}$, and consequently the probability of resolving a crossover with shrinkage with $p^{(2)}_{shrink} = 1 - p^{(2)}_{sev}$.

Figure \ref{fig5_sevVSshrink_V3}B shows the three distinct ways in which the newly-created crossover at $d$ can be resolved by shrinkage or severing: 1) the microtubule shrinks (is severed) without reaching $2d$, 2) the microtubule shrinks (is severed) after reaching $2d$ but without being severed there, 3) the microtubule shrinks (is severed) after reaching $2d$ and after being severed there. The third case bears a dependency on $p^+$.

We first notice that $p^{(2)}_{shrink}$ can be split in two probabilities, i.e. $p^{(2)}_{shrink} = q_{sev} + q_{shrink}$, where $q_{sev}$ is the probability of shrinkage after severing at $2d$ (path $\longrightarrow \longrightarrow \swarrow$ of Figure \ref{fig5_sevVSshrink_V3}B), while $q_{shrink}$ is the probability of shrinkage without any severing (paths $\longrightarrow \downarrow$ or $\searrow$ of Figure \ref{fig5_sevVSshrink_V3}B). Furthermore, since $q_{sev}$ depends on the dynamic behaviour of the microtubule just after the severing event, it carries a dependency on $p^+$ and can be split again in $q_{sev} \left( p^+ \right) = p^+ q_{sev,+} + \left( 1 - p^+ \right) q_{sev,-}$, where $q_{sev,\sigma}$ is the probability of shrinkage after being severed at $2d$ with the newly-created plus end in the state $\sigma$. The derivation of $q_{sev, \sigma}$ can be found in Appendix B.

As regards the probability $q_{shrink}$ that microtubules shrink below $d$ without being severed there, we observe that such a probability accounts all cases in which crossovers are resolved by shrinkage in absence of the crossover at $2d$, i.e. $p^{(1)}_{shrink}$, except for those cases in which microtubules that in principle would have shrunk back, do not have enough time to do so because they are severed at $2d$. We denote this probability with $q_{ns}$, and hence $q_{shrink} = p^{(1)}_{shrink} - q_{ns}$. The derivation of $q_{ns}$ can be found in Appendix B.

Therefore, the final expressions for the probabilities of resolving a crossover with a severing and with a shrinkage are
\begin{equation}       \label{p2shrink_final}
p^{(2)}_{shrink} \left( p^+ \right) = p^{(1)}_{shrink} - q_{ns} + q_{sev} \left( p^+ \right),
\end{equation}
\begin{equation}       \label{p2sev_final}
p^{(2)}_{sev} \left( p^+ \right) = p^{(1)}_{sev} + q_{ns} - q_{sev} \left( p^+ \right).
\end{equation}
Notice that $q_{sev} \left( p^+ \right)$ can be rewritten as
$$
q_{sev} \left( p^+ \right) = q_{sev,-} - \left( q_{sev,-} - q_{sev,+} \right) p^+,
$$
where the term in the braces is always positive. Indeed, since a microtubule initially in the growing state takes more time to completely depolymerize than a microtubule in the shrinking state, its probability of resolving the crossover at $d$ before being severed there is smaller than in the opposite case. Consequently, from Eq. (\ref{p2sev_final}) the probability $p^{(2)}_{sev} \left( p^+ \right)$ of resolving a crossover with a severing event grows linearly with $p^+$. 

By making use of this two-crossovers theory, we finally give a new estimate of the critical probability of rescue-after-severing by calculating the probabilities $p_{cr} \left( p^+ \right)$ and $\frac{1}{N} \sum\limits_{i=1}^N  \left\langle 1 - \delta_{c_i,0} \right\rangle$ of Eq. (\ref{amplification-condition_final}). In order to do that, we first define $p_{2d}$ as the probability to have a severing event at $2d$ before an eventual severing event at $d$. The derivation of $p_{2d}$ can be found in the Appendix B.

We now define the three events $A$, $B$, and $C$ as
\begin{align*}
A &= \mbox{shrinkage of microtubule below $d$ after severing at $2d$}, \\
B &= \mbox{severing event at $2d$ before an eventual severing event at $d$},\\
C &= \mbox{severing event at either $d$ or $2d$}.
\end{align*}
The three events are nested as $A \subset B \subset C$, and their probabilities are $\mathbb{P} \left( A \right) = q_{sev} \left( p^+ \right)$, $\mathbb{P} \left( B \right) = p_{2d}$, and $\mathbb{P} \left( C \right) = p_{sev}^{(2)} \left( p^+ \right) - q_{sev} \left( p^+ \right) + p_{2d} = p_{sev}^{(1)} - q_{ns} + p_{2d}$. Thus, it holds
\begin{equation}\begin{split}     \label{pcr}
p_{cr} \left( p^+ \right) &= \mathbb{P} \left( A | B \right) = \frac{\mathbb{P} \left( A \cap B \right)}{\mathbb{P} \left( B \right)} \\
& = \frac{\mathbb{P} \left( A \right)}{\mathbb{P} \left( B \right)} = \frac{q_{sev} \left( p^+ \right)}{p_{2d}},
\end{split}\end{equation}
and
\begin{equation}\begin{split}     \label{average-delta-sum}
\frac{1}{N} \sum_{i=1}^N  \left\langle 1 - \delta_{c_i,0} \right\rangle &= \mathbb{P} \left( B | C \right) = \frac{\mathbb{P} \left( B \cap C \right)}{\mathbb{P} \left( C \right)} \\
& = \frac{\mathbb{P} \left( B \right)}{\mathbb{P} \left( C \right)} = \frac{p_{2d}}{p_{sev}^{(1)} - q_{ns} + p_{2d}}.
\end{split}\end{equation}
Plugging $p_{cr}$ and $\frac{1}{N} \sum_{i=1}^N  \left\langle 1 - \delta_{c_i,0} \right\rangle$ into the equality associated to inequality (\ref{amplification-condition_final}), we finally calculate the critical threshold for the probability of rescue-after-severing $p^+_{crit,(2)}$, that is
\begin{equation}\begin{split}               \label{p+_crit_final}
p^+_{crit,(2)} = & \, \frac{ 1 }{ 2 S \Delta q_{sev} \beta \left( 1 - R_{d}^- \left( d \right) \right) } \\ 
& \times \bigg\{ \left( M^+_0 - S \alpha \beta \right)\left( 1 -R_{d}^- \left( d \right) \right) - S  \Delta q_{sev} \beta  R_{d}^- \left( d \right) \bigg. \\
& \quad \bigg. + \sqrt{ \left[ \left( M^+_0 - S \alpha \beta \right)\left( 1 - R_{d}^- \left( d \right) \right) + S \Delta q_{sev} \beta R_{d}^- \left( d \right) \right]^2 - 4 S \Delta q_{sev} \beta \left( 1 - R_{d}^- \left( d \right) \right) } \bigg\},
\end{split}\end{equation}
where
$$
\alpha = p_{sev}^{(1)} + q_{ns} - q_{sev,-},
$$
$$
\beta = \frac{1}{p_{sev}^{(1)} - q_{ns} + p_{2d}},
$$
$$
\Delta q_{sev} = q_{sev,-} - q_{sev,+}.
$$

Table \ref{table2-crit-p+} shows a very good agreement between our predicted critical probability of rescue-after-severing in the two-crossovers approximation and the critical probability obtained with our simulations in the whole grid of transverse microtubules for different choices of dynamic parameters, confirming our hypothesis that, in order to study the critical properties of the system, we can approximate the entire grid of transverse microtubules with just two of them without any considerable loss of accuracy.

\begin{table}
  \begin{center}
  \begin{tabular}{c c c c | c c c c c}
    \hline
    \hline
    \multicolumn{4}{c|}{\multirow{2}{*}{Dynamic parameters}} & Critical point & Critical point & Relative error & Critical point & Relative error \\
    & & & & (simulations) & (1-cross. theory) & & (2-cross. theory)  \\
    \hline
    $v^+$ & $v^-$  & $r_c$  & $r_r$ & $p_{crit}^+$& $p_{crit,(1)}^+$ & $\Delta p^+_{(1)}$ & $p_{crit,(2)}^+$ & $\Delta p^+_{(2)}$ \\
    $\mu\mbox{m} \, \mbox{s}^{-1}$ & $\mu\mbox{m} \, \mbox{s}^{-1}$ & $\mbox{s}^{-1}$ & $\mbox{s}^{-1}$ & - & - & - & - & - \\
    \hline 
    0.10 & 0.250 & 0.020 & 0.020 & 0.360 & 0.316 & 0.122 & 0.361 & 0.003 \\ 
    0.08 & 0.275 & 0.016 & 0.022 & 0.338 & 0.297 & 0.121 & 0.337 & 0.003 \\
    0.15 & 0.225 & 0.020 & 0.020 & 0.142 & 0.108 & 0.239 & 0.144 & 0.014 \\
    0.10 & 0.250 & 0.030 & 0.015 & 0.882 & 0.819 & 0.071 & 0.864 & 0.020 \\
    0.10 & 0.250 & 0.015 & 0.030 & 0.089 & 0.068 & 0.236 & 0.103 & 0.157 \\
    0.10 & 0.250 & 0.030 & 0.015 & 0.800 & 0.733 & 0.084 & 0.780 & 0.025 \\
    0.10 & 0.275 & 0.020 & 0.030 & 0.285 & 0.240 & 0.158 & 0.285 & 0.000 \\
    0.10 & 0.250 & 0.010 & 0.020 & 0.054 & 0.041 & 0.241 & 0.066 & 0.222 \\
    0.08 & 0.225 & 0.015 & 0.025 & 0.208 & 0.179 & 0.139 & 0.213 & 0.024 \\
    0.12 & 0.225 & 0.020 & 0.025 & 0.175 & 0.140 & 0.200 & 0.179 & 0.023 \\
    0.08 & 0.250 & 0.002 & 0.020 & 0.510 & 0.455 & 0.108 & 0.497 & 0.025 \\
    \hline
    \hline
  \end{tabular}
  \caption{Comparison $p^+_{crit}$ vs $p_{crit,(1)}$ vs $p_{crit,(2)}$ for different sets of dynamic parameters as $v^+$, $v^-$, $r_c$, and $r_r$. All other model parameters are those of Table \ref{table1-parameters}. $\Delta p^+_{(1)} = \frac{ p_{crit}^+ - p_{crit,(1)}^+}{ p_{crit}^+}$ and $\Delta p^+_{(2)} =\frac{ p_{crit}^+ - p_{crit,(2)}^+}{ p_{crit}^+}$ represent the relative error of the one and two-crossovers theory to the computationally measured critical value for the probability of rescue-after-severing.}
  \label{table2-crit-p+}
  \end{center}
\end{table}

\section{Discussion}

Our aim was to obtain a deeper insight into the conditions under which templated severing of microtubules at microtubule crossovers can lead to exponential proliferation of a new population of microtubules, as observed in the recent experiments on the light-induced reorientation of the plant microtubule cortical array. To that end we separately considered the role of the microtubule growth state, be it bounded or unbounded, and that of the rescue-after-severing effect previously identified as a key component of the amplification process. Simulations revealed a striking difference between the unbounded and the bounded microtubule growth regimes. In the unbounded-growth regime, which appears to be salient for the experimental situation, amplification due to templated severing will occur even in the absence of rescue-after-severing. The reason is that in this growth regime microtubules in principle have infinite lifetime, allowing them (and their descendants after severing) to be severed without limit, which by itself is sufficient to drive the amplification. There still is a role for the probability of rescue-after-severing, but only as a moderator for rate of amplification and the probability of success per microtubule. In contrast, in the bounded-growth regime an microtubule can in principle only be severed a finite number of times. In this case amplification can only occur if the process is biased by a sufficiently high probability of rescue-after-severing. When the system is below a critical value of this parameter, a newly nucleated microtubule, and all of its descendants through severing, is sure to go extinct. The value of this critical rescue-after-severing probability depends strongly on the probability of a newly-severed microtubule to cross the interval between neighboring transverse microtubules, so that it can be severed in turn, a crucial step in the amplification process. This prompted us to develop a novel approach to calculating the appropriate first passage time distribution, using an approach that may find application in other stochastic systems as well. This formed the basis of approximate calculation of the critical rescue-after-severing probability, which compares favourably with the results obtained from simulations. 

While our work sheds light on the initial phase of the amplification process, understanding the later stages and the stability of the final state remains a challenging problem. Here we have neglected a number of important effects. First, the transverse microtubules were taken to be inert, while in reality they are also dynamic and will tend to be broken down over time as more and more of the available tubulin is incorporated into the exponentially growing population of longitudinal microtubules. This will remove opportunities for severing, and therefore tend to dampen the amplification again.  Moreover, as the amplification process develops, the availability of free tubulin dimers, which surely are a limited resource in the cell, is also bound to decrease, which in turn affects both the growth dynamics and nucleation rate. Given our results here, the first effect, depression of the growth speed, could in fact switch the microtubules from the unbounded to the bounded-growth regime, which likely decelerates the amplification process. We are currently exploring these issue, which will be the subject of a follow-up paper.

\begin{acknowledgments}
The work of MS was supported by the ERC 2013 Synergy Grant MODELCELL. The work of BMM is part of the research program of the Dutch Research Council (NWO).
\end{acknowledgments}

\section{Appendix A: main features of the Dogterom-Leibler model}

\subsection{Splitting probabilities in the interstitial strip}

If a microtubule plus end impinges on a transverse microtubule, it creates a crossover. After the creation of the crossover the plus end is located at $x \in \left( nd, (n+1)d \right)$, and, as long as this condition is fulfilled, the dynamics of microtubules is described by the Dogterom-Leibler model for microtubules with their minus end at $nd$, regardless the occurrence of a severing event. Without any loss of generality, we can set $n=0$ and length $l=x$. 

Due to the dynamic instability of the plus end, the microtubule either reaches length $x=d$ or shrinks back to length $x=0$. The occurrence probability of either of these events is described by the so-called \textit{splitting probabilities} $R^\sigma_0 \left( x \right)$ and $R^\sigma_d \left( x \right)$, that describe the probability that a microtubule with initial state $\sigma$ and initial length $x$ arrives first at length $0$ or $d$ respectively. Conservation of probability implies that $R^\sigma_0 \left( x \right) + R^\sigma_d \left( x \right) = 1$.

It is possible to show \cite{Mulder2012MicrotubulesTimes} that
\begin{equation}         \label{splitting-xtod+}
R_{d}^+ \left( x \right) = \frac{ e^{x/\overline{l}} - \frac{r_r v^+}{r_c v^-} }{ e^{d/\overline{l}} - \frac{r_r v^+}{r_c v^-} },
\end{equation}
\begin{equation}         \label{splitting-xtod-}
R_{d}^- \left( x \right) = \frac{ \frac{r_r v^+}{r_c v^-}  \left( e^{x/\overline{l}} - 1 \right) }{ e^{d/\overline{l}} - \frac{r_r v^+}{r_c v^-} },
\end{equation}
\begin{equation}         \label{splitting-xto0+}
R_{0}^+ \left( x \right) = \frac{ e^{d/\overline{l}} - e^{x/\overline{l}} }{ e^{d/\overline{l}} - \frac{r_r v^+}{r_c v^-} },
\end{equation}
and
\begin{equation}         \label{splitting-xto0-}
R_{0}^- \left( x \right) = \frac{ e^{d/\overline{l}} - \frac{r_r v^+}{r_c v^-} e^{x/\overline{l}}  }{ e^{d/\overline{l}} - \frac{r_r v^+}{r_c v^-} }.
\end{equation}
Interestingly, these expressions hold for both bounded and unbounded-growth case. This is a direct consequence of the fact that in a strip both regimes produce a steady-state solution \cite{Govindan2004SteadyGeometry}.

\subsection{Microtubule lifetime and survival probability}

The lifetime density function $L_\sigma\left( t | x \right)$ of a microtubule with initial length $x$ and initial state $\sigma$, is defined as the distribution of the time needed by microtubules to completely depolymerize.

In the bounded-growth regime all microtubules have a finite lifetime, hence $L_\sigma \left( t | x \right)$ is normalized to $1$. However, in the unbounded-growth a fraction of microtubules grows linearly in time. It follows that for unbounded-growth microtubules, the lifetime density function can be defined only for the fraction of microtubules the lifetime of which is finite.

In the bounded-growth regime, the lifetime density functions are \cite{Bicout1997}
\begin{equation}\begin{split}      \label{lifetime-distribution_+}
L_+ \left( t | x \right) = & \ \Theta\left( t - \frac{x}{v^-} \right) \frac{r_c}{v^+ t + x} \, e^{-\left[ r_r \left( v^+ t + x \right) + r_c \left( v^- t - x \right) \right]} \\
& \times \Bigg[ x \, I_0 \left( \frac{2}{v^+ + v^-} \sqrt{r_r r_c \left( v^+ t + x \right) \left( v^- t - x \right)} \right) \big. \\
& \qquad \Bigg. + \frac{v^+}{r_c} \sqrt{\frac{r_c \left( v^- t - x \right)}{r_r \left( v^+ t + x \right) }} I_1 \left( \frac{2}{v^+ + v^-} \sqrt{r_r r_c \left( v^+ t + x \right) \left( v^- t - x \right)} \right)\Bigg],
\end{split}\end{equation}
\begin{equation}\begin{split}      \label{lifetime-distribution_-}
L_- \left( t | x \right) = & \ \delta \left( t - \frac{x}{v^-} \right) \, e^{-r_r t} + \Theta\left( t - \frac{x}{v^-} \right)   \sqrt{ \frac{r_c r_r}{ \left( v^+ t + x \right) \left( v^- t - x \right) } } \, x \\
& \times e^{-\left[ r_r \left( v^+ t + x \right) + r_c \left( v^- t - x \right) \right]} I_1 \left( \frac{2}{v^+ + v^-} \sqrt{r_r r_c \left( v^+ t + x \right) \left( v^- t - x \right)} \right),
\end{split}\end{equation}
where $I_0 \left( \cdot \right)$ and $I_1 \left( \cdot \right)$ are the modified Bessel functions of order $0$ and $1$, respectively. 

In order to obtain the densities in the unbounded-growth regime, we need to by divide Eq. (\ref{lifetime-distribution_+}) and Eq. (\ref{lifetime-distribution_-}) by $ 1 - S_+ \left( \infty | x \right)$ and $1 - S_- \left( \infty | x \right)$ respectively, where $S_\sigma \left( \infty | x \right)$ is the fraction of microtubules with initial length $x$ and initial state $\sigma$ that never completely depolymerize. Due to their finite lifetime, in the bounded-growth regime these fractions are identically $0$. The fractions $S_\sigma \left( \infty | x \right)$ are called \textit{ultimate survival probabilities}, and they are
\begin{equation}              \label{ultimate-survival-prob_+}
S_+ \left( \infty | x \right) =
\begin{cases} 
  1 - \frac{r_c v^-}{r_r v^+} \, \exp\left[- \frac{r_r v^+ - r_c v^-}{v^+ v^-} x\right] & \qquad \mbox{if unbounded-growth regime},\\
  0 & \qquad \mbox{if bounded-growth regime},
\end{cases}
\end{equation} 
\begin{equation}              \label{ultimate-survival-prob_-}
S_- \left( \infty | x \right) =
\begin{cases}
  1 - \exp\left[- \frac{r_r v^+ - r_c v^-}{v^+ v^-} x\right] & \qquad \mbox{if unbounded-growth regime},\\
  0 & \qquad \mbox{if bounded-growth regime}.
\end{cases}
\end{equation}

\section{Appendix B: derivation of the size of the offspring of a microtubule}

Here, we derive the expression for the size of the offspring of a microtubule in the one and two-crossovers approximations.

\begin{figure}[htbp]   
  \centering
  \includegraphics[scale=0.27]{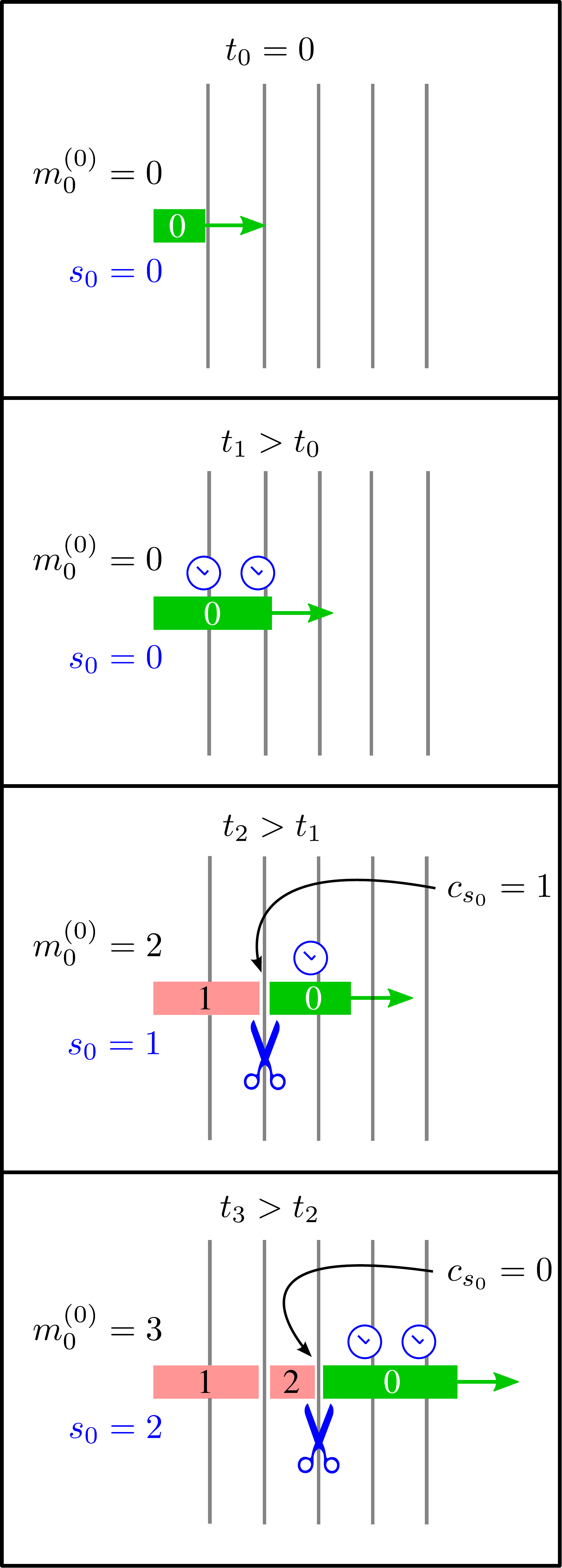}
  \caption{Schematic of the count of the size of the offspring $m_0^{(1)}$ of a microtubule labelled by $0$ created by severing in the growing state. When a crossover is created, the competition severing-shrinking takes place, and if the severing occurs, the counter for the number of severing events $s_0$ gains one unity, whilst the size of the offspring gains $1 + c_{s_0}$. We keep track of the leading microtubule as it can generate other descendants, further increasing $m_0^{(1)}$. We do not keep track of the lagging microtubules created by severing.}  \label{fig3_Mplus-count}
\end{figure}

First, we introduce the one-crossover approximation by removing the dependency on $p^+$ from the r.h.s. of Eq. (\ref{m_i}). We assume that $b_{c_{j_i}} = 0$ for every $j_i$. This implies that, when a severing event occurs at $nd$, $n>1$, then all previous crossovers are resolved by a severing event. With this approximation, we replace $m_i$ with
\begin{equation}    \label{m_i^0}
m_i^{(1)} = s_i + \sum\limits_{j_i=1}^{s_i} c_{j_i},
\end{equation}
see Figure \ref{fig3_Mplus-count}. Analytically, we cannot calculate neither $s_i$ nor $c_{j_i}$, but these quantity are easily measurable with computer simulations. We average $m_i^{(1)}$ over $N = 10^5$ simulations to find the first approximation for $M^+$, i.e. $M^+_{(1)} = \frac{1}{N} \sum\limits_{i=1}^N m_i^{(1)} = 2.61$. Therefore, from Eq. (\ref{p+_crit}), we can calculate the first estimate of the critical probability of rescue-after-severing, i.e. $p^+_{crit,(1)} = 0.316$, against the computationally measured one $p^+_{crit} = 0.360$. Table \ref{table2-crit-p+} shows a comparison between $p^+_{crit,(1)}$ and $p^+_{crit}$ for different sets of dynamic parameters. The table shows that, even though our one-crossover approximation provides a reasonable estimate of the critical probability, we systematically underestimate it. 

Now, we introduce the two-crossovers approximation by assuming that after a severing at $nd$, $n>1$, all crossovers at $d$, $2d$, $\dots$, $\left( n - 2 \right) d$ are resolved by a severing event, while the crossover at $\left( n - 1 \right) d$ is resolved by a shrinkage with probability $p_{cr} \left( p^+ \right)$ and by a severing with probability $1 - p_{cr} \left( p^+ \right)$. Our aim is to give a better estimate of $p^+_{crit}$ then in the one-crossover approximation. Here
$$
b_{c_{j_i}} = \left( 1 - \delta_{c_{j_i},0} \right) p_{cr} \left( p^+ \right).
$$
With this definition for $b_{c_{j_i}}$, we approximate $m_i$ with
\begin{equation}    \label{m_1^1}
m_i^{(2)}  = s_i + \sum_{j_i=1}^{s_i} \Big[ c_{j_i} - \left( 1 - \delta_{c_{j_i},0} \right) p_{cr} \left( p^+ \right) \Big].
\end{equation}
From this equation, we can observe that
\begin{equation} \begin{split}    \label{1-delta_calc}
\sum_{j_i=1}^{s_i} \left( 1 - \delta_{c_{j_i},0} \right) p_{cr} \left( p^+ \right) & = p_{cr} \left( p^+ \right) \Big[ s_i - \left( \delta_{c_{1},0} + \delta_{c_{2},0} + \cdots + \delta_{c_{s_i},0} \right) \Big] \\
& = p_{cr} \left( p^+ \right) \Big[ s_i - s_i \left\langle \delta_{c_i,0} \right\rangle \Big] \\
& = p_{cr} \left( p^+ \right) s_i \left\langle 1 - \delta_{c_i,0} \right\rangle,
\end{split}\end{equation}
where the average value $\left\langle \delta_{c_i,0} \right\rangle$ is calculated over all severing events that a leading microtubule undergoes along its lifetime. Consequently, $\left\langle 1 - \delta_{c_i,0} \right\rangle$ is the fraction of severing events that a leading microtubule undergoes at $nd$ with $n>1$. If we combine Eqs. (\ref{m_i^0}), (\ref{m_1^1}), and (\ref{1-delta_calc}) together, and we average over $N$, we obtain
\begin{equation} \begin{split}    \label{M+1-app}
M^+_{(2)} & = \frac{1}{N} \sum_{i=1}^N m_i^{(2)} \\
& = \frac{1}{N} \sum_{i=1}^N \Big[ s_i + \sum_{j_i=1}^{s_i} c_{j_i} - p_{cr} \left( p^+ \right) s_i \left\langle 1 - \delta_{c_i,0} \right\rangle \Big] \\
& = M^+_{(1)} - p_{cr} \left( p^+ \right) S \frac{1}{N} \sum_{i=1}^N  \left\langle 1 - \delta_{c_i,0} \right\rangle,
\end{split} \end{equation}
where $S = \frac{1}{N} \sum\limits_{i=1}^N s_i$, and where we assumed that the correlation between the number of severing events that occur along the lifetime of a microtubule and the fraction of them that occur at $nd$ with $n>1$ is neglectable. In this case, if $N \gg 1$, for the law of large numbers $\frac{1}{N} \sum\limits_{i=1}^N  \left\langle 1 - \delta_{c_i,0} \right\rangle$ is the probability that a microtubule is severed at $nd$ with $n>1$, sampled over all cases in which a severing event has occurred. By replacing $M^+$ with $M^+_{(2)}$ in Eq. (\ref{amplification-condition_2}), we obtain the final amplification condition (\ref{amplification-condition_final}). Table \ref{table2-crit-p+} shows that the two-crossovers reproduces the computationally measured critical probability of rescue-after-severing with a good degree of accuracy.

\section{Appendix C: derivation of severing and shrinkage probabilities}

In order to calculate $q_{sev,\sigma}$ we first define the following random variables: 
\begin{equation}\begin{split}
&\tau_{d} = \mbox{severing waiting time at $d$}, \\
&\tau_{2d} = \mbox{severing waiting time at $2d$}, \\
&T_d = \mbox{FPT from the first to the second crossover, i.e. from $d$ to $2d$}, \\
&T_{\sigma} \left( x \right) = \mbox{lifetime of a microtubule with initial state $\sigma$ and initial length $x$}, \\
&\widetilde{\tau}_{2d} = \mbox{severing waiting time at $2d$ given that the severing occurs}.
\end{split}\end{equation}
$\tau_{d}$ and $\tau_{2d}$ have probability density function $W_{k \theta} \left( t \right)$ defined in Eq. (\ref{gamma_sever-wait_distr}), while the probability density function of $T_d$ is $F_{0d} \left( t \right)$ from Eq. (\ref{FPTD}). The probability density function of $T_{\sigma} \left( x \right)$ is $L_+ \left( t | x \right)$ defined in the Appendix A. Finally, the probability density function of $\widetilde{\tau}_{2d}$ can be calculated by observing that the event ``severing'' and the event ``shrinkage'' are independent. Therefore, the cumulative function $\Phi_{\widetilde{\tau}_{2d}} \left( t \right) = \mathbb{P} \left[ \widetilde{\tau}_{2d} < t \right]$ can be written as
\begin{equation}
\Phi_{\widetilde{\tau}_{2d}} \left( t \right) = \mathbb{P} \left[ \left( \tau_{2d} < t \right) \cap \left( \tau_{2d} < T_+ \left( 0 \right) \right) \right] = \frac{1}{Z_W} \int_0^{t} dt' \, W_{k,\theta} \left( t' \right) \int_{t'}^{\infty} dt'' \, L_+ \left( t'' | 0 \right),
\end{equation}
where
\begin{equation}
Z_W = \int_0^{\infty} dt \, W_{k,\theta} \left( t \right) \int_t^{\infty} dt' \, L_+ \left( t' | 0 \right).
\end{equation}
Thus, the probability density function $\widetilde{W}_{k\theta} \left( t \right)$ of $\widetilde{\tau}_{2d}$ is
\begin{equation}            \label{pdf_tildetau}
\widetilde{W}_{k\theta} \left( t \right) = \frac{d}{dt}\Phi_{\widetilde{\tau}_{2d}} \left( t \right) = \frac{1}{Z_W} W_{k\theta} \left( t \right) \int_t^{\infty} dt' \, L_+ \left( t' | 0 \right).
\end{equation}

The probability $q_{sev,\sigma}$ is the probability that a microtubule reaches $2d$, it is severed there with newly-created plus end in the state $\sigma$, and finally shrinks back below $d$ \textit{before} being severed at $d$. In $S_+ \left( \infty | 0 \right)$ of the cases (i.e. for indefinitely growing microtubules), this event occurs if $T_1 = \tau_d - T_d - \tau_{2d} - T_\sigma \left( d \right) > 0$, with probability density function of $T_1$ defined by
\begin{equation}
P_{ T_1 } \left( t \right) = \int_{\mathbb{R}^3} dt'  dt''  dt''' \, W_{k\theta} \left( t + t' + t'' + t''' \right) F_{0d} \left( t' \right) W_{k\theta} \left( t'' \right) L_\sigma \left( t''' | d \right).
\end{equation}
In the remaining $1 - S_+ \left( \infty | 0 \right)$ of the cases (i.e. for microtubules with a finite lifetime), the event occurs if $T_2 = \tau_d - T_d - \widetilde{\tau}_{2d} - T_\sigma \left( d \right) > 0$, and if it is severed at $2d$, i.e. if $\tau_d < T_+ \left( 0 \right)$. The probability density function of $T_2$ is
\begin{equation}
P_{ T_2 } \left( t \right) = \int_{\mathbb{R}^3} dt'  dt''  dt''' \, \widetilde{W}_{k\theta} \left( t + t' + t'' + t''' \right) F_{0d} \left( t' \right) W_{k\theta} \left( t'' \right) L_\sigma \left( t''' | d \right).
\end{equation} 
Hence, the final expression for $q_{sev,\sigma}$ is
\begin{equation} \begin{split}     \label{qsev_sigma}
q_{sev,\sigma} = R_{d}^+ \left( 0 \right) & \left\{ S_+ \left( \infty | 0 \right) \int_0^{\infty} dt \, P_{ T_1 } \left( t \right)  + \left[ 1 - S_+ \left( \infty | 0 \right) \right] \int_0^{\infty} dt \, P_{ \tau_d - T_0} \left( t \right) \int_0^{\infty} dt \, P_{ T_2 } \left( t \right) \right\}  \left[ 1 - S_+ \left( \infty | d \right) \right].
\end{split}\end{equation}

To calculate $q_{ns}$ we first define the random variable $\widetilde{T}_{+}$ as the time that a microtubule initially in the growing state and with plus end in $2d$ needs in order to return in the shrinking state at $2d$, given that no severing event occurs at $2d$. Similarly to the derivation of $\widetilde{W}_{k\theta} \left( t \right)$, we can derive the probability density function of $\widetilde{T}_{+}$, that is
\begin{equation}
\widetilde{L}_{+} \left( t \right) = \frac{1}{Z_L} L_+ \left( t | 0 \right) \int_0^t dt' \, W_{k,\theta} \left( t' \right),
\end{equation}
with
\begin{equation}
Z_L = \int_0^{\infty} dt \, L_+ \left( t | 0 \right) \int_0^t dt' \, W_{k,\theta} \left( t' \right).
\end{equation}

Therefore, as $q_{ns}$ is the probability that a microtubule reaches length $2d$ and would return to length $d$ but it cannot because it is severed at $2d$, the two conditions that our random variables have to fullfil are $\tau_{2d} < T_+ \left( 0 \right)$ and $\tau_d > T_d + \widetilde{T}_{+} + T_- \left( d \right)$. The former condition had already been discussed before, whilst the latter is associated to the probability density function
\begin{equation}
P_{\tau_d - T_d - \widetilde{T}_{+} - T_- \left( d \right)} \left( t \right) = \int_{\mathbb{R}^3} dt' dt'' dt''' \, W_{k\theta} \left( t + t' + t'' + t''' \right) F_{0d} \left( t' \right) \widetilde{L}_{+} \left( t'' \right) L_- \left( t''' | d \right).
\end{equation}
Therefore
\begin{equation}    \label{qns}
q_{ns} = R_{0d}^+ \left[ 1 - S_+ \left( \infty | 0 \right) \right] \int_{-\infty}^0 dt \, P_{\tau_d - T_+ \left( 0 \right)} \left( t \right) \int_0^{\infty} dt \, P_{\tau_d - T_d - \widetilde{T}_{+} - T_- \left( d \right)} \left( t \right).
\end{equation}

Finally, in order to calculate the probability $p_{2d}$ to have a severing event at $2d$ before an eventual severing event at $d$, we notice that we have two different cases. In the first case, the microtubule reaches length $2d$ and it is severed there before being severed at $d$, i.e. $\tau_d > T_d + \tau_{2d}$. In the second case, the microtubule reaches $2d$ and it is severed there before being severed at $d$, i.e. $\tau_d > T_d + \widetilde{\tau}_{2d}$, given that the event ``severing'' wins the competition against the event ``shrinkage'' at $2d$, or $\tau_{2d} < T_+ \left( 0 \right)$. The probability density functions associated to these conditions are, respectively,
\begin{equation}
P_{\tau_d - T_d - \tau_{2d}} \left( t \right) = \int_{\mathbb{R}^2} dt' dt'' \, W_{k\theta} \left( t + t' + t'' \right) F_{0d} \left( t' \right) W_{k\theta} \left( t'' \right),
\end{equation}
\begin{equation}
P_{\tau_d - T_d - \widetilde{\tau}_{2d}} \left( t \right) = \int_{\mathbb{R}^2} dt' dt'' \, W_{k\theta} \left( t + t' + t'' \right) F_{0d} \left( t' \right) \widetilde{W}_{k\theta} \left( t'' \right),
\end{equation}
\begin{equation}
P_{\tau_{2d} - T_+\left( 0 \right)} \left( t \right) = P_{\tau_{d} - T_+\left( 0 \right)} \left( t \right).
\end{equation}
Then
\begin{equation}\begin{split}       \label{pd}
p_{2d} = R_{0d}^+ & \left\{ S_+ \left( \infty | 0 \right) \int_0^{\infty} dt \, P_{\tau_d - T_d - \tau_{2d}} \left( t \right) + \left[ 1 - S_+ \left( \infty | 0 \right) \right] \int_{-\infty}^0 dt \, P_{\tau_{2d} - T_+\left( 0 \right)} \left( t \right) \int_0^{\infty} dt \, P_{\tau_d - T_d - \widetilde{\tau}_{2d}} \left( t \right) \right\}.
\end{split}\end{equation}

%\bibliographystyle{unsrt}
%\bibliography{severing}

\end{document}